# Scale Collapse of Vortices at Porous/Fluid Interfaces

Justin Courter, Vishal Srikanth, Thibaut Kemayo, and Andrey V. Kuznetsov*

Department of Mechanical and Aerospace Engineering, North Carolina State University, Raleigh, NC 27695, USA

The interaction between externally generated turbulence and porous media is central to many engineering and environmental flows, yet the fate of macroscale vortical structures at a porous/fluid interface remains uncharacterized. By numerically simulating the turbulent flow using a two-dimensional (2D) direct numerical simulation (DNS), we investigate the penetration, breakdown, and turbulent kinetic energy (TKE) transport of macroscale vortices impinging on porous matrices with high porosities $\phi$ = 0.80-0.95. For all porosities considered, macroscale vortices collapse abruptly at the porous interface and do not persist within the matrix, supporting the pore-scale prevalence of turbulence even under strong external forcing. Although vortex impingement injects TKE into the porous medium through turbulent transport at the interface, this supplied TKE is rapidly redistributed and dissipated as the flow reorganizes to satisfy pore-scale geometric constraints. Deeper within the porous layer, turbulence is sustained primarily by local shear production associated with pore-scale velocity gradients, and the internal flow becomes increasingly independent of upstream conditions. Variations in porosity modulate the balance between production and dissipation of TKE by altering geometric confinement and the local flow conditions within the pores although the dominant turbulent length scale within the porous matrix remains set by the pore size. These results demonstrate that porous media act as a robust geometric filter that enforces pore-scale-dominated turbulence regardless of the external forcing.

---

## 1 Introduction

Scale separation of turbulent vortices within and surrounding porous media represents a frontier in the understanding of how turbulent structures – spanning from the macroscale size to the microscale size – are transported across porous/fluid interfaces. Since porous media are characterized by a solid matrix permeated by interconnected voids [1], these multiscale turbulence interactions are frequently encountered in practice across a wide range of natural and engineered flow environments.

Experimental evidence for turbulence within porous media is well established; Mayer [2] found a sharp increase in the pressure loss in a porous layer when increasing the Reynolds number which was attributed to the presence of turbulence (see Fig. 3.7 in [2]). Early debate centered on whether geometric constraints imposed by solid obstacles would suppress macroscale turbulent structures. It was found that eddy size is limited to the distance between pores [3], supporting the view that macroscale turbulence cannot survive within the matrix. This principle is known as the pore scale prevalence hypothesis (PSPH). Uth *et*

* Email address for correspondence: avkuznet@ncsu.edu

*al.* [4] and Jin & Kuznetsov [5] confirmed the PSPH for different geometries using Direct Numerical Simulation (DNS) studies in infinitely spanning periodic porous media. Additional studies [6–8] further solidified the view that porous media inherently limit the time and length scale of turbulence via pore-size constraints which suppress the development of macroscale vortices within the matrix. Within this regime, turbulence manifests as microscale vortices that form downstream of individual pores [9] and are eventually advected into the inter-pore free stream where they dissipate.

An exception in turbulent scale suppression behavior in porous media may exist at porosities exceeding roughly 98% in which the flow constraints are sufficiently nonlimiting that macroscale structures are allowed to exist [10]. However, it is important to note that this high-porosity exception was observed in idealized periodic porous media, a configuration that may inherently prevent the initial formation of macroscale turbulent structures. This leaves a critical, complementary question open: if macroscale turbulence is instead produced outside the porous medium and forced into it, would it then survive? Thus, the characterization of externally produced macroscale vortices that subsequently impinge on a porous layer is an important canonical flow configuration.

The impact of externally generated turbulence on porous media has been studied by Suga *et al.* [11] and Manes *et al.* [12] who investigated the impact of wall permeability on turbulence. However, most work emphasizes a partially porous channel flows in which the porous/fluid interface is oriented along the streamwise direction and intralayer motion remains laminar. As such, the internal turbulent kinetic energy (TKE) transport within porous media to strong external vortical forcing remains poorly characterized.

The interaction of macroscale turbulent flow in porous media spans several highly impactful heat and mass transfer applications in areas such as electronics cooling [13], industrial thermal management [14], filtration [15], oil extraction from porous surfaces [16], wildfire modeling [17], and aerodynamic performance optimization [18]. While several researchers have either studied microscale turbulence in periodically repeating porous media [19] or used macroscale turbulence modeling approaches to study porous/fluid interfaces, the microscale resolved turbulence interactions at porous/fluid interfaces has remained underexplored despite its crucial relevance to these practical engineering applications.

Given the practical relevance of vortex impingement on porous layers, several experimental studies have been performed in recent years to investigate and characterize the interaction between externally generated impinging rings and porous layers. Musta & Krueger [20] studied the interaction of a vortex ring impinging on multiple permeable screens at porosities up to $\phi = 0.84$ and Reynolds numbers up to 3000, finding that lower porosities break up the original vortex ring into smaller vortical structures and significantly impact the decay rate of total energy. Xu *et al.* [21] presented an experimental study on the impact of Reynolds number and porosity on synthetic vortex jets impinging on a porous wall. At low Reynolds number and porosity ($\phi = 0.30$), small-scale vortex rings formed downstream of individual pores. As Reynolds number increased, Kelvin-Helmholtz instability intensified due to higher velocity gradients within the pore channels, promoting vorticity cancellation and the formation of a new vortex ring. Lin *et al.* [22] found that the porosity of concave hemispherical cavities impacts the flow behavior downstream of the cavity. At low porosities ($\phi = 0.10$ and 0.22), vortex rings impinging on a porous cavity are broken into incoherent, small-scale structures, but at larger porosities ($\phi = 0.40$ and 0.56), the cavity merges vorticities and creates spatial coherence larger than the original vortex ring.

Together, these studies suggest that porosity governs both the survival of macroscale vortices in porous media and with it, kinetic energy transport. However, the focus of these previous studies was limited to single-layer porous meshes rather than thick, porous media composed of many layers - the geometry used in this study. A central unanswered question remains: how do macroscale vortices propagate through a porous layer consisting of numerous solid obstacles? Do the observations of vortex persistence in single layer studies continue to be valid, or does the cumulative impact of successive porous obstacles completely break down the macroscale vortex? The fate of macroscale vortices impinging at a porous/fluid interface remains unresolved and motivates the present study.

The present study addresses both gaps in understanding the vortex dynamics at the clear-fluid/porous interface and the resulting internal TKE development within the porous matrix by employing a 2D DNS framework to simulate the turbulent wake generated by a single solid obstacle impinging on a downstream porous layer across porosities $\phi$ = 0.80-0.95. We divide this interaction into two distinct phases: (1) impingement at the porous/fluid interface - that is, the plane where a clear-fluid region (devoid of all porous obstacles) and a porous matrix meet and (2) the subsequent internal development of the flow within the porous matrix. The full TKE budget is evaluated to elucidate the dynamics of vortex penetration and energy transfer. The remainder of the paper is organized as follows: §2 discusses the computational setup and methods, §3 looks at the collapse of the macroscale vortex upon impingement on the porous layer along with the TKE dynamics, and §4 summarizes findings and proposes additional topics of interest.

## 2 Computational Setup and Methods

### *2.1 Physical model and computational domain*

The simulation geometry consists of a square bluff body of side length *D* placed upstream of an explicitly resolved in-line porous matrix composed of square solid obstacles. The purpose of the bluff body is to generate macroscale vortical structures in the clear-fluid region so that their interaction with the porous/fluid interface can be examined. A companion case without the upstream bluff body is also simulated. In that case, the flow approaching the porous layer is not externally forced by a wake, and the simulation therefore provides a baseline for isolating the TKE generated by the porous matrix itself. The porous layer occupies a finite streamwise section of the domain, beginning at the porous/fluid interface and extending through the explicitly resolved matrix. This configuration allows the incoming macroscale wake to interact with several rows of pore-scale obstacles before exiting into the downstream clear-fluid region.

The simulation domain is two-dimensional and has dimensions $15D$ in the streamwise direction and $6D$ in the transverse direction. A uniform flow velocity, $u_{in}$, of 1m/s is prescribed at the inlet, the fluid density $\rho$ is 1 kg/m$^3$, and the viscosity, $\mu$ is 0.001 Pa×s, and $D$ is 10 m, such that the Reynolds number at the bluff body, $Re_D$, is 10,000 where,

$$Re_D = \frac{\rho \times u_{in} \times D}{\mu} \tag{1}$$

and a gauge pressure of 0 Pa is specified at the outlet. Periodic boundary conditions are imposed at the upper and lower boundaries to simulate an infinitely repeating porous layer in the transverse direction, since we are primarily interested in the impingement of vortices normal to the porous/fluid interface. Both the square bluff body and porous layer solid obstacle surfaces satisfy the no-slip condition.

The porous matrix is defined by the pore pitch $d$ and the square-obstacle side length $s$. For the in-line array of square obstacles simulated in this paper, the porosity can be calculated as,

$$\phi = 1 - \left(\frac{s}{d}\right)^2 \quad \textbf{(2)}$$

In the present work, the porosity is varied by changing the obstacle size $s$ while keeping the pitch ($d$), the solid obstacle shape, and simulation domain dimensions fixed. This approach is favored over changing the pitch ($d$) since changing the pitch will change the scale separation between the bluff body and the pore size significantly at high porosities such that the size of the macroscale vortex will then approach that of the separation distance between the porous layer solid obstacles. However, we note that the study can also be performed by simultaneously increasing the pore pitch and the bluff body size when changing the porosity. The porosities considered are $\phi$ = 0.80, 0.85, 0.90, and 0.95. This range was selected to test high-porosity media, where macroscale structures are most likely to survive if the matrix can support them.

Two sampling regions are used throughout the analysis. The *impingement region* corresponds to the portion of the porous layer where the bluff body wake directly impinges on the porous/fluid interface, approximately $7D \leq x \leq 8D$ (the porous matrix) and $2.5D \leq y \leq 3.5D$ (directly downstream of the bluff body). The *non-impingement region* corresponds to neighboring portions of the same porous layer, $7D \leq x \leq 8D$ with $y < 2.5D$ or $y > 3.5D$ (not downstream of the bluff body). This distinction is used to organize our analysis of the results based on the exposure to the incoming macroscale vortex, identifying the influence of direct vortex impingement rather than changes in the matrix geometry.

Additional methodological details, including an assessment of whether the velocity fluctuations are turbulent and stochastic, justification for the use of a 2D rather than 3D DNS framework, descriptions of the numerical procedure, sampling strategy, and TKE metrics, grid and temporal-resolution studies, and benchmarking against prior DNS data, are provided in the Supplementary Material.

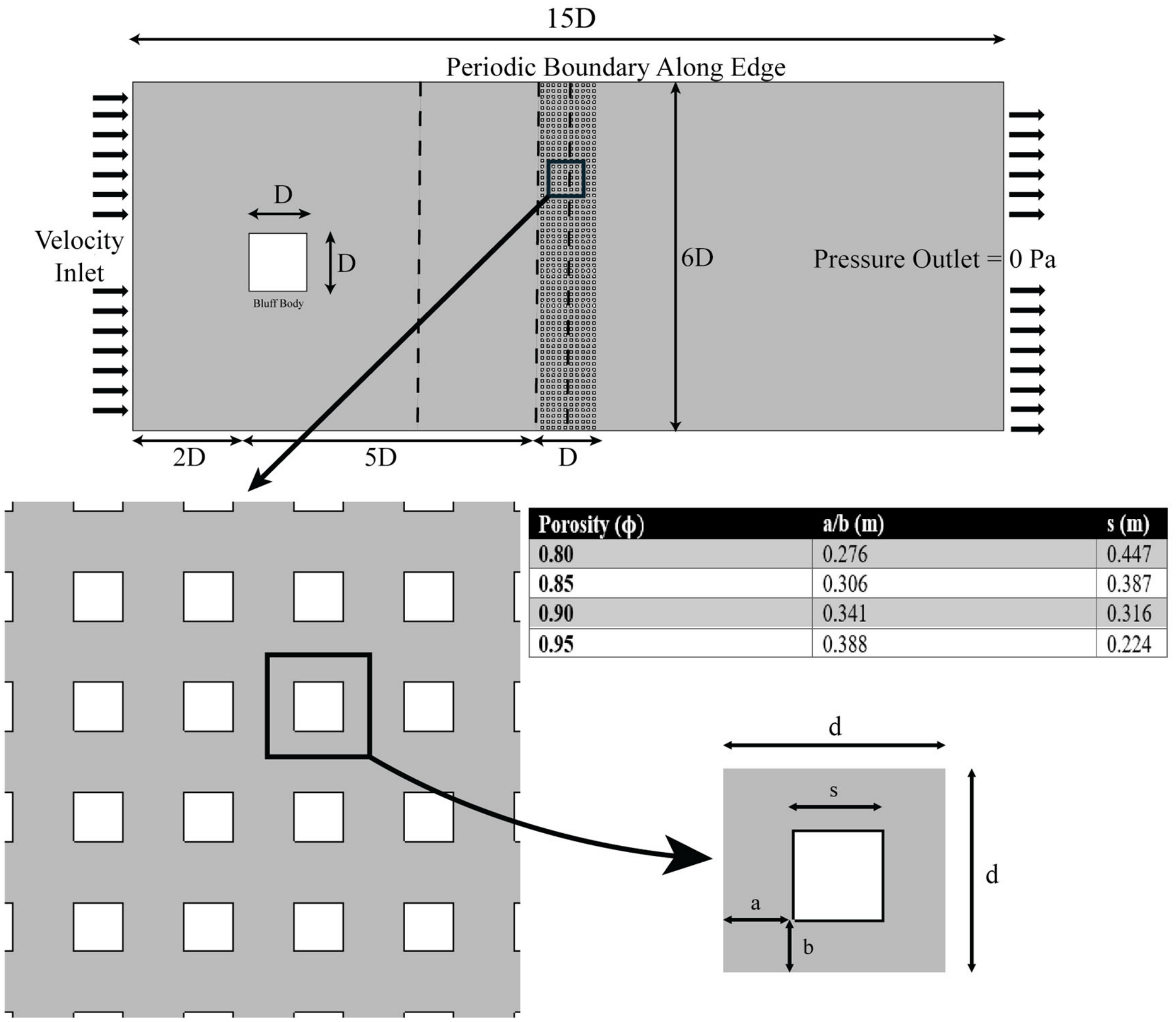

| Porosity (ϕ) | a/b (m) | s (m) |
|---|---|---|
| 0.80 | 0.276 | 0.447 |
| 0.85 | 0.306 | 0.387 |
| 0.90 | 0.341 | 0.316 |
| 0.95 | 0.388 | 0.224 |

Figure 1: The representative geometry of size $15D\,x\,6D$ with uniform flow at the inlet, a square bluff body of side $D$ placed $2D$ from the inlet, and the porous domain placed $5D$ downstream from the bluff body. The dashed lines (from left to right) correspond to the locations where the correlation functions are computed to find the turbulent length scale: (i) midway between the bluff body and porous matrix, (ii) at the clear fluid/porous interface, and (iii) within the porous matrix.

## 3 Results and Discussion

The goal of this study is to understand the transport mechanism of macroscale turbulent vortices when they are externally forced perpendicular to a porous/clear fluid interface. To characterize this fundamental phenomenon, we have organized the discussion as follows:

i. First, we investigate the dynamics of how a macroscale vortex is advected to the porous/fluid interface and whether the solid obstacles of the porous matrix break down the macroscale vortex to microscale size. We will address the question of macroscale vortex survival in a porous medium and the existence of an entrance length for vortex dissipation and establish a baseline of the flow behavior.

ii. Next, we investigate the effect of porosity on the macroscale vortex transport into the porous layer, specifically targeting porous media with higher porosities (0.80-0.95) to study the boundary cases where macroscale vortices are most likely to persist.
iii. Finally, we analyze the transport of turbulent kinetic energy inside the porous medium by comparing regions with and without macroscale vortex impingement to highlight the interfacial fluid mechanics.

### *3.1 Characterization of macroscale vortex impingement on a microscale porous layer*

Macroscale vortical structures are introduced onto porous media by upstream turbulence generation from both active mechanisms like pumping and fan-assisted convection or passive mechanisms like bluff bodies or wind gusts. This production of macroscale vortices introduces large scale separation of vortical structures with the porous medium, which leads to multiscale interaction at the porous/fluid interface. To study this phenomenon, we simulated 2D turbulent flow around a macroscale bluff body with a porous layer of porosity 0.80 situated downstream of it. Under this geometric configuration, large vortices are generated behind the bluff body that are turbulent in nature ($Re_D$ = 10,000) and their length scale is similar to the size of the bluff body – creating an order of magnitude scale separation between the macroscale vortex and the microscale pores. These macroscale vortices are advected downstream by the flow and impinge on the porous matrix without any significant diminishment of their length scale (Figure 2a).

While these macroscale vortical structures persist upstream of the porous layer, we observed virtually no macroscale structures either within or downstream of the porous layer at a porosity of 0.80. Further examination of the flow distribution within the porous layer confirmed that the macroscale vortex is abruptly diminished at the porous/fluid interface indicating geometric confinement due to enhanced local shear near the surfaces of the solid obstacles that comprise the porous matrix. Within the porous layer, the turbulent length scale is therefore confined to the microscale level as evidenced by the microscale vortices that are formed behind the individual solid obstacles of the porous matrix (Figure 2c). These microscale vortices represent the starting point of turbulence production within the porous layer. Additionally, consistent with Srikanth & Kuznetsov [9] and Agnaou *et al.* [23], we observed the formation of secondary oscillatory instability of the flow streamlines deep inside the porous layer (Figure S4 in the Supplementary Material), which is a characteristic of fully developed turbulent flow at these porosities.

At the exit of the porous layer, jets of turbulent flow emerge from the individual pores and interact in the clear-fluid region to form vortical structures. However, despite the macroscale size of these downstream vortical structures, they bear no resemblance to the macroscale vortices that were initially produced by the bluff body. Taken together, the streamline plots visualize a clear sequence: macroscale vortices formed upstream of the porous matrix collapse to pore-scale motions inside, then re-expand into larger-scale structures behind the porous layer.

To quantitatively evaluate the length scale of the coherent structures in the model, we calculated the two-point velocity fluctuation correlation function upstream of the porous layer (see §S.3 in the Supplementary Material for further information), at the porous/fluid interface, and at the center of the porous layer (Figure 2b). The model predicted that, while the width of the correlation at and upstream of the porous/fluid interface is of macroscale size, the width of the individual peaks of the correlation function at the center of the porous layer is of microscale size. This confirmed that the widths of the correlation function upstream of the porous layer are identical, indicating that the dominant vortex length scale at these locations is set by the bluff body size. In contrast, the rapid decay in the correlation function within the porous layer suggests pore-scale suppression where the length scales of velocity fluctuations within the porous matrix are significantly smaller than those in the surrounding clear-fluid region.

Intriguingly, the correlation function at the center of the porous layer exhibits a large-scale correlation superimposed on the individual peaks. To understand this behavior, we analyzed the budget of turbulent kinetic energy transport by interpreting the results within the Kraichnan-Leith-Batchelor framework of 2D turbulence, as detailed in §S.2 of the Supplementary Material. The distribution of TKE within the porous layer reveals that the initially elevated peak value of TKE at the porous/fluid interface drops twofold within the first three columns of solid obstacles leading to weak fluctuations in the flow (Figure 2d). This is caused by significant negative production in the stagnant face of the porous/fluid interface – a characteristic of porous flows [6] – and a high dissipation rate of the flow at the interface that is caused by strong shear at the solid obstacle surface (Figure 2f). As a result of this weak turbulence near the center of the porous layer, correlations persist between the jetting flow patterns between the solid obstacles of the porous matrix. Thus, these correlations do not represent a single macroscale vortex surviving inside the porous layer, but rather the effect of the undamped momentum deficit introduced by the macroscale vortex wake (see Figure S5 in the Supplementary Material).

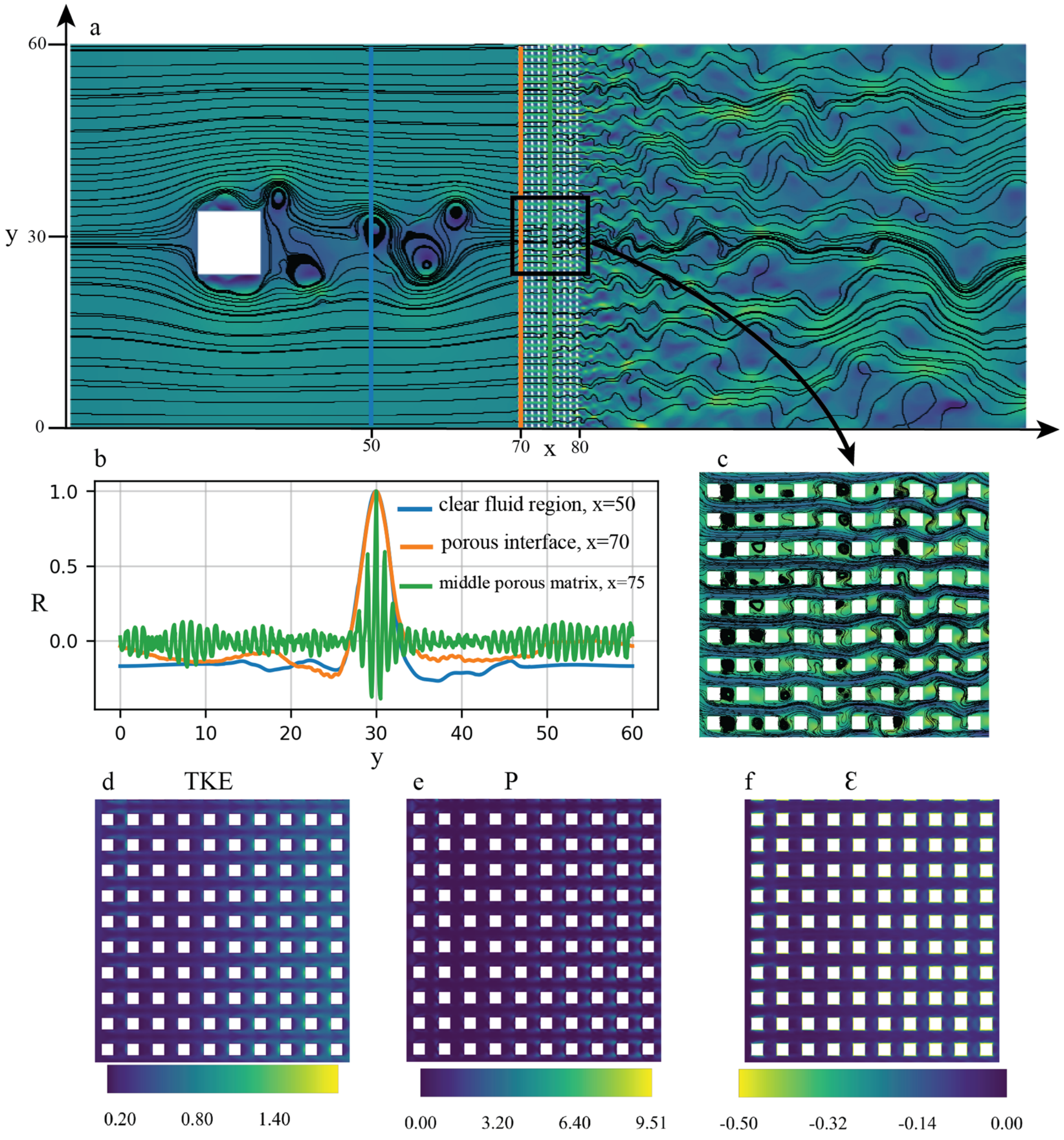

Figure 2: Macroscale vortices transition to microscale size when they impinge on a porous layer ($\phi$= 0.80): (a) Instantaneous $x$-velocity contours and streamlines show macroscale vortex production by the bluff body and its transport to the porous layer. (b) Two-point correlation function of velocity fluctuation reveals the diminished correlation value at the center of the porous layer (green) when compared to the upstream (blue) and interface (orange) locations signifying vortex breakdown. (c) A zoomed view of the porous layer at the interface confirms only microscale vortex persistence in the porous layer. The distributions of time-averaged (d) TKE, (e) TKE production: $P$, and (f) TKE dissipation: $\varepsilon$; these terms dominate the TKE transport budget and reveal the sequence of microscale transport within the porous layer. **Note**: (d), (e), and (f) exhibit the same domain as (c).

### *3.2 The effect of porosity on macroscale vortex destruction at the porous/fluid interface*

Our simulation results demonstrated that macroscale vortices are rapidly broken down upon impingement at a porous layer with constrained porosity. To provide the strongest confirmation that the macroscale vortices do not persist inside porous layers, we investigated whether this behavior is consistent across different porosities by focusing on high porosity media, and we hypothesized that the large void volume in these domains would allow the macroscale vortex to be transported deeper into the porous layer. We repeated our simulation analysis by increasing the porosity of the porous layer (0.80, 0.85, 0.90, 0.95), ensuring a constant scale separation between the bluff body ($D$) and the pore scale ($d$). Similar to the porosity of $\phi = 0.80$, the macroscale vortex is completely broken down at the porous/fluid interface at all tested values of porosity. As the porosity increases, the shear layers that are formed between the solid obstacles in the microscale flow distribution are characterized by lower values of shear stress. Therefore, the formation of shedding microscale vortices and the lack of a strong shear layer envelope around the microscale pore wake leads to strong instabilities in the microscale flow. However, this observation is also independent of the impinging macroscale vortex since these instabilities are observed in both the impingement and non-impingement regions of the porous layer, as discussed later.

The primary influence of the impinging macroscale vortex on the microscale flow in the porous layer becomes evident at high porosity at the porous/fluid interface, where the localized macroscale flow direction is not aligned with the direction of the free-stream velocity (the $x$-direction). This results in macroscale streamline tortuosity at the interface with a non-trivial penetration depth between 2-5 pore sizes depending on the porosity (see Figure S5 in Supplementary Material). This behavior is more readily visualized when observing macroscale vortex collapse as it impinges upon the porous matrix (Figure 3). As the macroscale vortex approaches the porous/fluid interface, it introduces a region of low pressure and high vorticity near the porous obstacles that generates macroscale flow and influences the microscale vortex formation at a shallow penetration distance into the pores. This is characterized by strong streamline curvature and asymmetrical flow patterns at the microscale level around the solid obstacles of the porous matrix. The intense shear generated by this flow is associated with enhanced turbulent production and dissipation rate, which locally removes TKE from the flow and contributes to the breakdown of the macroscale vortex. In contrast, regions that are not exposed to vortex impingement exhibit more homogenous flow velocity distributions, which are similar to that of flows in periodic porous media. This homogeneity promotes streamwise turbulence transport and more symmetrical flow patterns. Deep inside the porous layer, the flow patterns become qualitatively indistinguishable between the impingement and non-impingement regions.

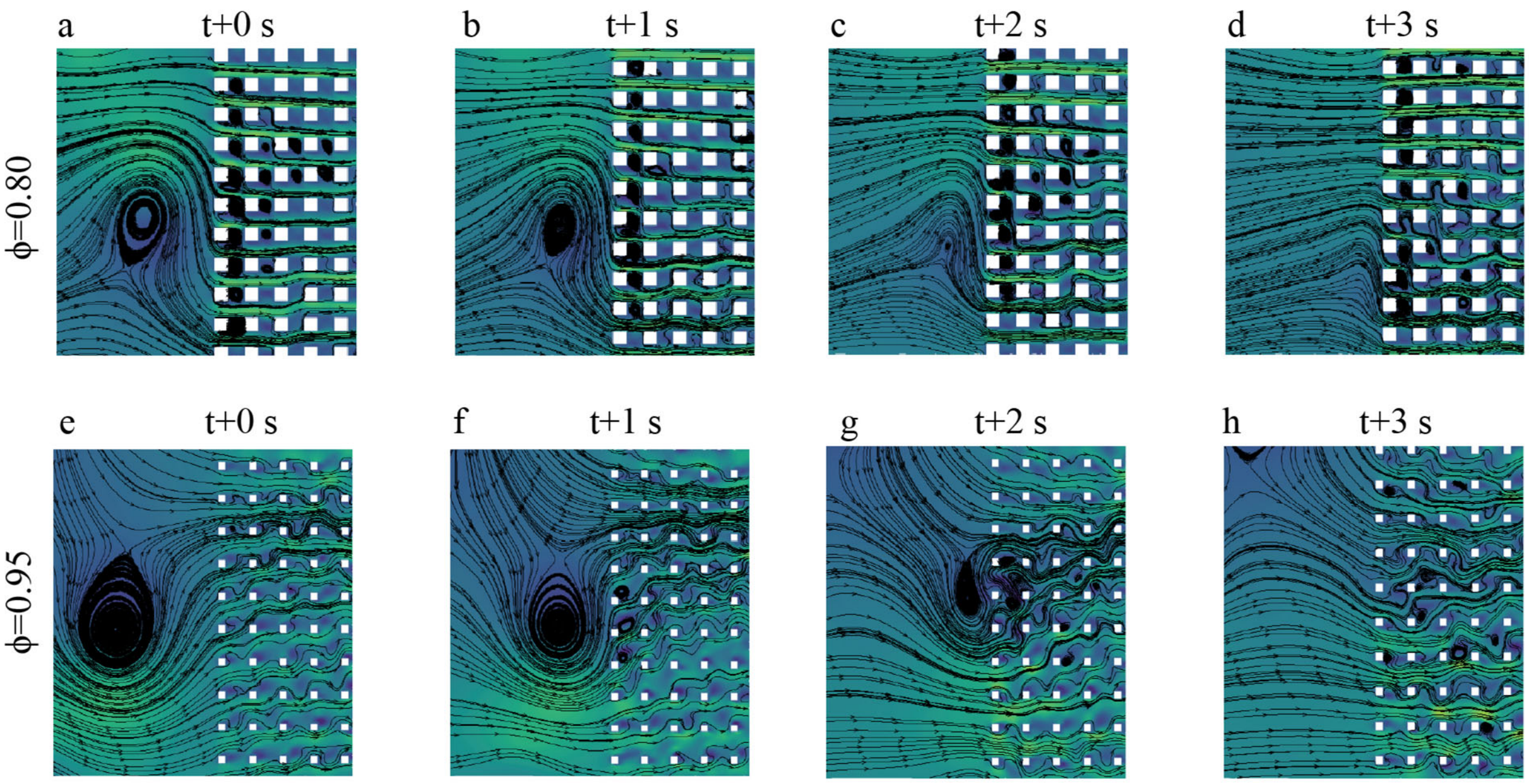

Figure 3: Contours of $x$-velocity and streamlines as a function of time for porosities $\phi = 0.80$ (a-d) and $\phi = 0.95$ (e-h). The macroscale vortex is clearly formed in (a, e) and moves toward the porous matrix in (b, f) where streamlines begin to bend and velocity gradients increase. It starts to dissipate as it nears the porous matrix (c, g) and in (d, h), the macroscale turbulent structure has lost its coherence.

Since the macroscale vortex influenced microscale flow patterns within the pores at high porosity, we asked whether this constituted a macroscale coherent structure within the porous layer. Therefore, we calculated the two-point correlation function of the velocity fluctuations at the same locations as previously: upstream of the porous layer, at the clear-fluid/porous interface, and at the center of the porous layer. The plots of the two-point correlation function and their associated integral length scales (area under the correlation function curve before the first zero crossing) can be seen in Figure 4 and Table S2 in the Supplementary Material, respectively, for all porosities analyzed. The correlation functions upstream and at the porous/fluid interface for different porosities confirmed nearly identical coherent length scales of the macroscale vortex prior to impingement. Interestingly, within the porous layer, the correlation function exhibited a tighter peak at higher porosities compared to the lower porosities, signaling a complete destruction of any macroscale coherence of the flow inside the porous layer despite the deeper penetrating influence of the macroscale vortex.

To understand this counterintuitive result, that deeper propagation of macroscale vortical motion did not result in a wider correlation function at the center of the porous layer, we examined the volume-averaged turbulent kinetic energy distribution inside the porous layer (Figure 5). At the porous/fluid interface, porous layers with higher porosities generally exhibited higher local volume-averaged TKE. This trend is reversed at the exit of the porous layer as the flow fully develops and microscale TKE is dominantly produced by the porous matrix. Note that the same trends are observed in regions where no vortex impingement occurs suggesting that this is an intrinsic property of the porous layer independent of macroscale influence. However, at higher porosities, the less constrained pore space causes a shorter entrance length for turbulence development. The rapid onset of pore-scale instabilities leads to high turbulent dispersion near the interface, effectively perturbing the induced macroscale flow patterns and destroying their spatial coherence. This rapid turbulent dispersion provides a physical explanation for the narrower correlation peaks observed at high porosities.

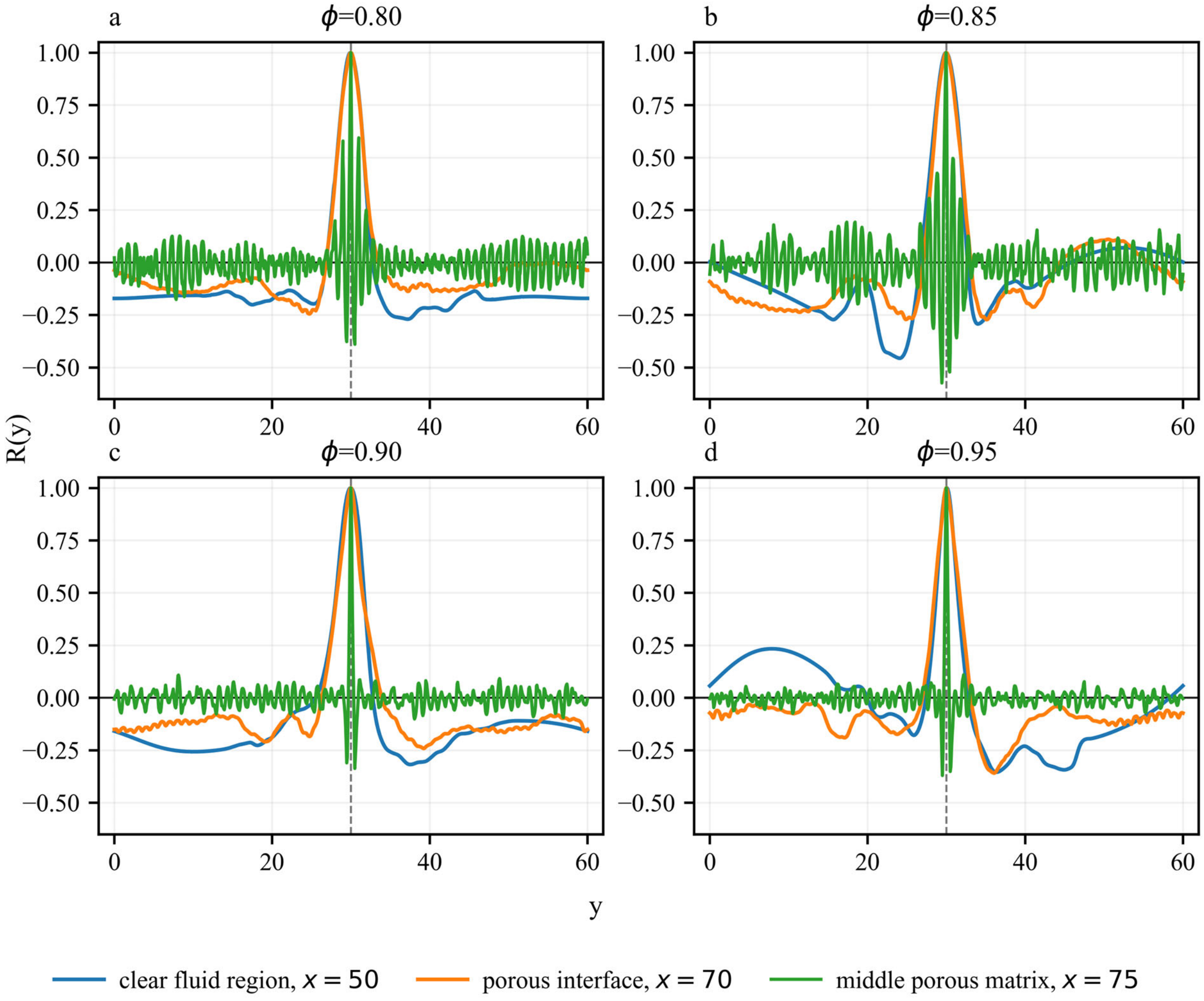

Figure 4: Correlation functions for porosities (a) 0.80, (b) 0.85, (c) 0.90, and (d) 0.95 for three regions of interest within the fluid domain (see Figure 1 for the exact sampling locations and Table S2 in the Supplementary Material for the integral length scales of each porosity). The first (blue) is the clear-fluid region directly downstream of the large solid obstacle where macroscale turbulence is unconstrained. Turbulent structures travel to the clear-fluid/porous interface (orange) after the clear-fluid region. At this location, the macroscale eddies continue to exist because they are not yet confined by the porous matrix. Finally, midway through the porous matrix (green), the eddy size is significantly diminished due to the pore constraints. The length scales in the impingement and impingement-free regions are similar, indicating that the macroscale vortices are confined to the porous length scale.

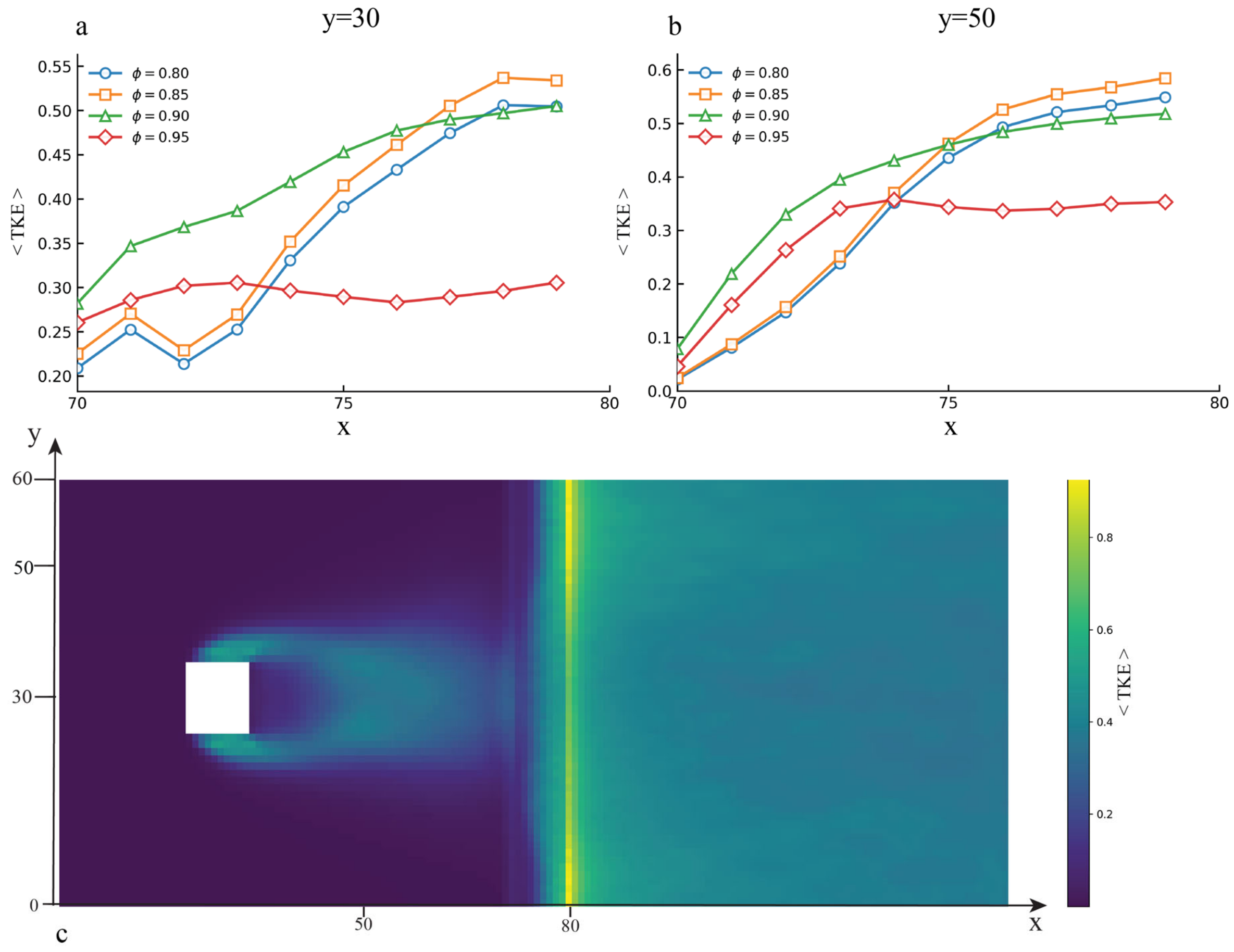

Figure 5: Volume-averaged TKE vs distance into the porous matrix for different porosities in (a) the region subjected to vortex impingement, at (b) the region free from the impact of vortex impingement, and (c), the volume-averaged TKE for the entire flow domain for $\phi = 0.80$.

### *3.3 The effect of macroscale vortex impingement on TKE transport inside the pores*

Based on our analysis, macroscale vortex impingement dramatically influences the TKE distribution inside the porous layer, especially close to the porous/fluid interface where the impingement occurs. We developed a more complete phenomenological understanding of these behaviors by calculating a budget of the TKE transport within the pores with five principal terms: production, pressure diffusion, viscous diffusion, turbulent transport, and viscous dissipation. The spatial evolution of TKE within the porous layer is governed by a competitive balance between inherited upstream fluctuations and locally generated pore-scale shear. By examining the TKE budget across varying porosities and impingement conditions, specifically comparing regions of direct vortex impingement with non-impinged and no-bluff body control cases, we delineate the mechanisms through which the porous matrix modifies the flow physics.

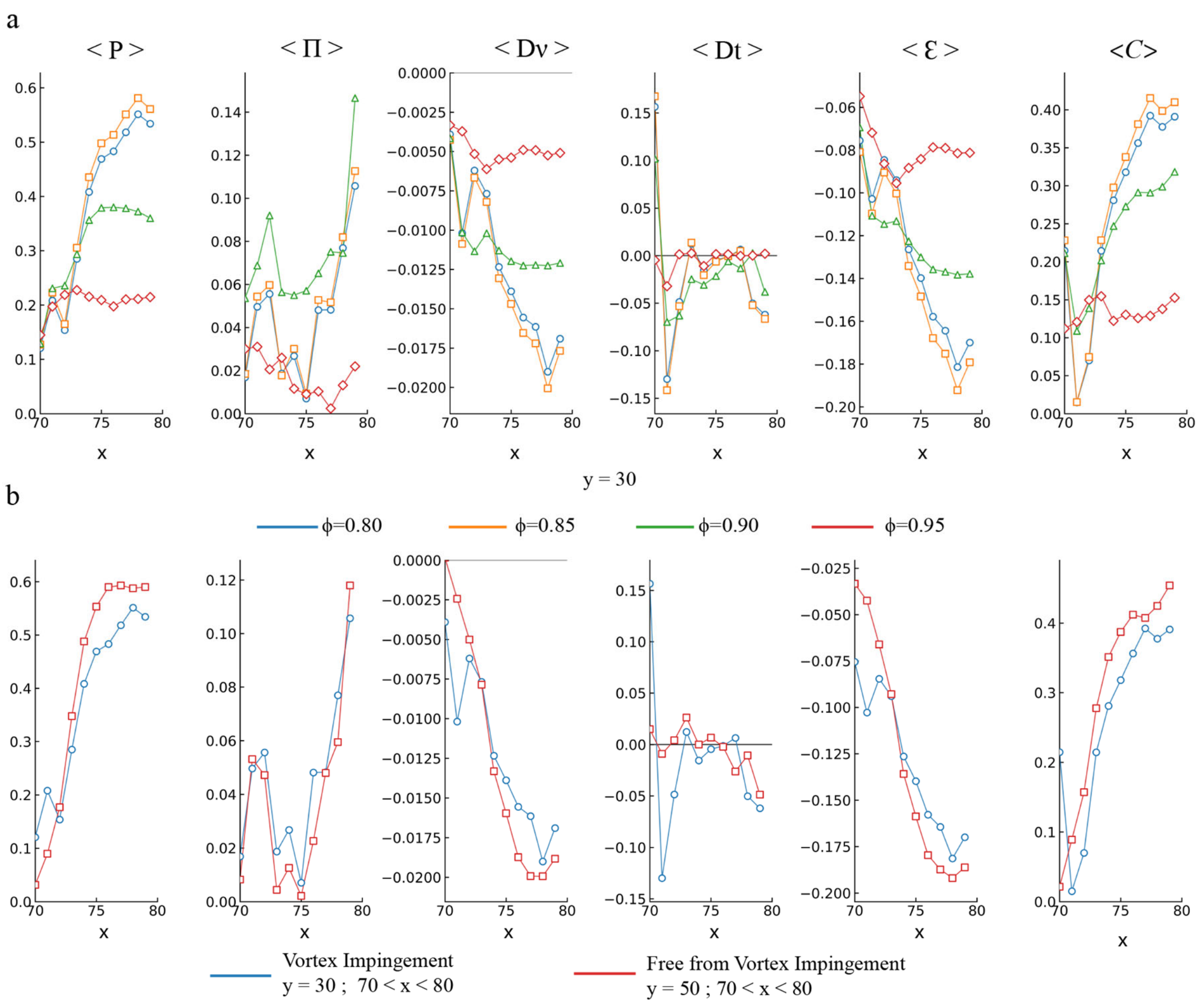

Figure 6: Terms in the volume-averaged TKE transport equation – from left to right: turbulent production ⟨P⟩, pressure diffusion ⟨Π⟩, viscous diffusion $\langle D_v \rangle$, turbulent transport $\langle D_t \rangle$, dissipation $\langle \varepsilon \rangle$, and total convection $\langle C \rangle$ which is the sum of the preceding terms at (a) the vortex-impinged region for different porosities ($y = 30;\ 70 < x < 80$) and at (b) the vortex-impinged region ($y = 30\ ; 70 < x < 80$) and the that free from vortex impingement ($y = 50\ ; 70 < x < 80$) for ϕ = 0.85.

The spatial evolution of the TKE budget confirms the streamwise transition of the flow from macroscale structural filtering of impinged vortices to the subsequent emergence of microscale vortices produced at the pore scale. At the porous interface, the initial TKE distribution is bifurcated between the impingement and non-impingement regions. In the vortex impingement region, the budget is dominated by TKE inflow into the porous layer characterized by high turbulent transport of TKE, which is the divergence of the triple velocity correlation. This is evidenced by a sharp positive peak in the transport term at the porous/fluid interface (Figure 6a), signifying a net influx of TKE. Conversely, the inflow remains only weakly turbulent or laminar (TKE < 0.1) in the non-impinged regions and the no-bluff body control case (Figure 7) resulting in negligible turbulent transport of TKE at the porous/fluid interface (Figure 6b, red line).

While this interfacial disparity highlights the immediate impact of the macroscale vortex, it also sets up a fundamental hydrodynamic trade-off that governs TKE production deeper within the matrix. Although the impingement region benefits from a substantial initial injection of turbulent energy, the macroscale wake inherently carries a momentum deficit. This deficit dampens the local mean velocity gradients inside the pores, thereby limiting the flow's ability to generate new microscale TKE through shear production. In stark contrast, the non-impingement regions receive unobstructed, higher-momentum flow, even though it lacks the initial TKE injection. This higher flow momentum preserves sharper velocity gradients across the solid obstacles, which ultimately provides a much higher capacity for local shear production once pore-scale instabilities are triggered deeper inside the porous matrix.

The TKE of the macroscale vortex in the impinged regions is rapidly attenuated, as indicated by negative turbulent transport of TKE following the first obstacle row. Physically, this transition represents a structural bottleneck where the porous geometry disperses the impinged macroscale vortices and diminishing their length scale (Figure 6) such that the decrease in the macroscale TKE occurs faster than local production mechanisms can sustain them. This budgetary redistribution causes TKE levels to drop after the inlet, especially at low porosity where the flow is most constrained (Figure 5a). Therefore, deep inside the porous layer (~2-5 obstacle layers), turbulence transport transitions from the external influence of the macroscale vortex to the internal generation of microscale vortices in the porous matrix.

Following this initial entrance region, the TKE budget reflects a localized equilibrium between production and viscous dissipation, which are the dominant terms of TKE transport, followed by pressure diffusion. TKE production is governed by the magnitude of the mean velocity gradients and Reynolds stresses. Pressure diffusion of TKE becomes significant due to the constant interaction of the flow with the solid obstacles of the porous matrix, creating oscillatory regions of stagnation pressure on the obstacle surfaces. However, because production is the predominant term in the TKE transport equation, the magnitudes of the mean velocity gradients and Reynolds stresses are strong candidates for explaining how TKE varies with porosity. The magnitude of production is constrained by the geometric confinement of the pores; as porosity decreases, the fluid is forced through narrower interstitial gaps, intensifying the mean velocity gradients between the high-speed interstitial gap centerlines and the no-slip obstacle walls. This intensification amplifies the shear production term, which acts as the primary driver of TKE growth.

One notable exception occurs for $\phi = 0.85$ where the production slightly exceeds that of $\phi = 0.80$; this behavior results from the larger Reynolds stresses in the $\phi = 0.85$ case, which increase the magnitude of the production term through their interaction with the mean velocity gradients. For all cases, dissipation is also enhanced in regions of large velocity gradients. Physically, sharper velocity gradients mean that neighboring fluid layers slide and stretch past one another more strongly, so viscosity does more work against the fluctuating motion, thereby acting as a sink in the TKE budget. As mean shear strengthens, the resulting turbulent fluctuations develop steeper gradients at smaller scales, leading to higher dissipation. As a result, volume-averaged production and dissipation tend to scale with one another and exhibit similar spatial trends (Figure 6)Figure 6: pores with large magnitude of production coincide with large magnitude of dissipation rate.

In lower-porosity cases, this stronger production allows the TKE to reach a larger peak farther downstream into the porous matrix than in higher-porosity cases (Figure 5a/Figure 5b). For higher porosities, production and dissipation attain their maximum values nearer to the entrance of the porous matrix and then reach a quasi-steady state with slow growth and/or decay (Figure 6a: $\phi = 0.90/0.95$). The streamlines are less constrained than in lower-porosity cases and can traverse multiple rows of obstacles; as a result, the mean velocity gradients are weaker, and the production peak is smaller, which limits the peak TKE.

Interestingly, non-impinged regions achieve higher peak production deep within the matrix than impinged regions (compare Figure 5a and Figure 5b; Figure 6b). This paradox arises because the upstream bluff-body wake introduces a momentum deficit, weakening mean velocity gradients and suppressing local shear production in the impinged flow (Figure 6b). In contrast, the unobstructed, higher-momentum flow preserves sharper velocity gradients, driving stronger TKE growth. TKE growth eventually saturates as transverse momentum exchange between obstacle wakes and interstitial channels diffuses local shear gradients. Figure 7 illustrates this for the no-bluff-body case and the vortex-impinged and impingement-free regions downstream of a bluff body at $\phi = 0.85$: deep within the matrix (x≈77), TKE profiles for all regions converge to a uniform state (0.5≤TKE≤0.6) despite different boundary conditions at the porous interface (TKE≈0.2 at the interface of the vortex-impinged region vs. TKE≈0 at the interface of the non-impingement region), confirming that internal porous geometry erases upstream flow history.

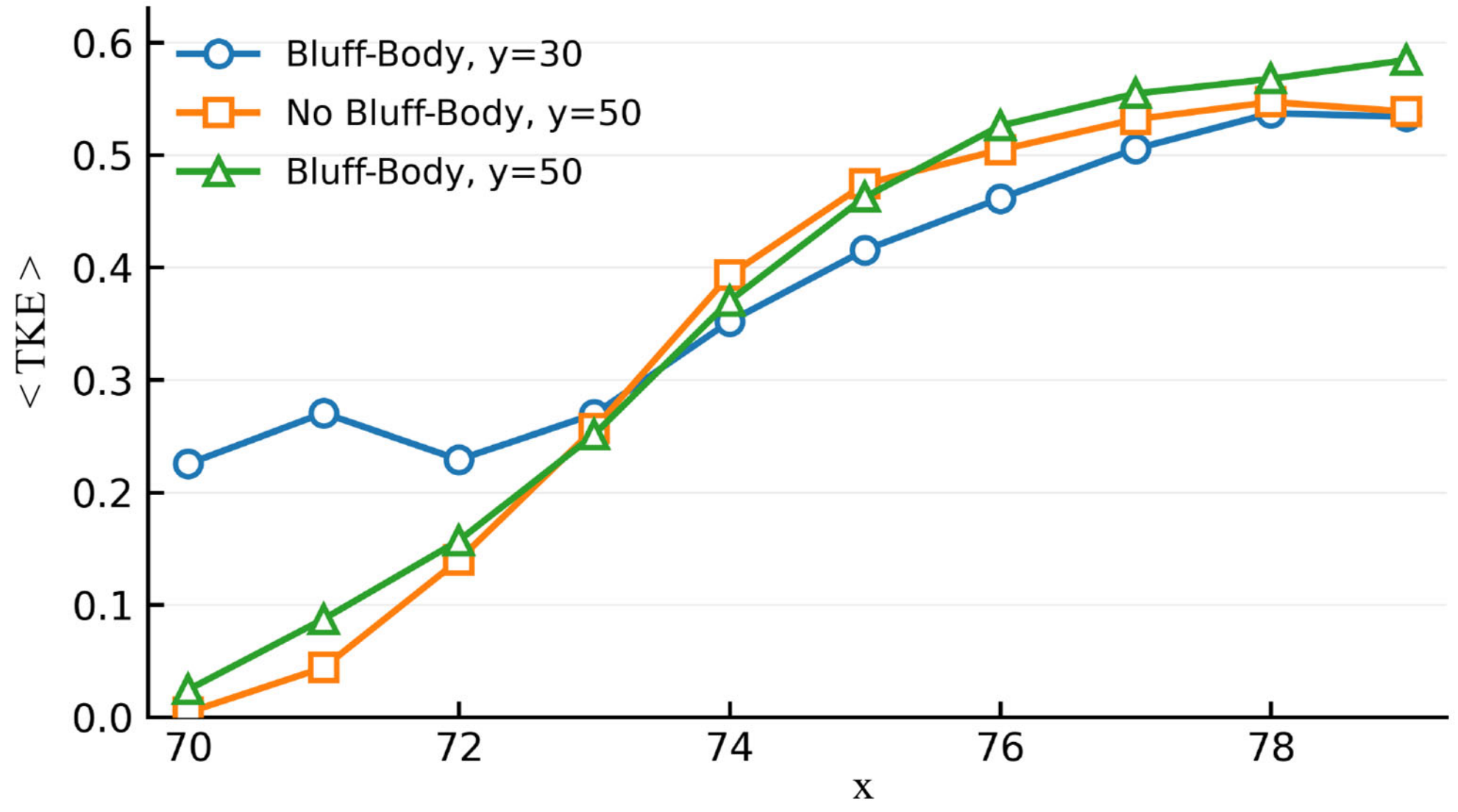

Figure 7: TKE comparison for a $\phi = 0.85$ with and without a large bluff body upstream of the porous matrix.

## 4 Conclusions

The interaction of macroscale turbulent structures impinging at porous/fluid interfaces and their evolution within porous media remains a critical challenge in fluid mechanics. In this paper, we investigated the impact of macroscale vortex impingement on the transport of TKE within a porous matrix considering different porosities (ϕ = 0.80, 0.85, 0.90, 0.95). We found that the macroscale vortices are broken down immediately upon impingement and this is simultaneously accompanied by microscale vortex production inside the porous matrix.

This abrupt scale collapse has direct consequences for the turbulent kinetic energy budget within the porous layer. Analysis of the volume-averaged TKE transport equation revealed that, although macroscale vortex impingement introduces a localized influx of turbulent energy at the porous/fluid interface, this externally influenced TKE is rapidly redistributed and dissipated as the flow reorganizes to satisfy pore-scale constraints. Deeper within the matrix, turbulence is sustained primarily by local shear stress-induced TKE production associated with the mean, transverse, microscale velocity gradients imposed by the solid obstacles, rather than by direct transport of macroscale structures. As a result, the internal turbulence becomes increasingly independent of upstream forcing as it moves deeper into the porous layer, approaching a locally determined equilibrium governed by the pore-scale geometry.

Porosity was found to play a dual role by simultaneously controlling geometric confinement and the TKE budget within the pores. Decreasing porosity intensified mean velocity gradients and tended to enhance both production and dissipation of TKE, leading to an overall inverse relationship between volume-averaged TKE and porosity. An intermediate porosity ($\phi = 0.85$) produced the highest turbulence levels due to a favorable balance between strong shear and sufficient pore connectivity. In contrast, higher-porosity matrices permitted deeper penetration of impingement-induced disturbances but exhibited weaker shear production, limiting the growth of TKE. Importantly, regions subjected to vortex impingement consistently displayed lower peak TKE production within the matrix than non-impinged regions, owing to the momentum deficit inherited from the upstream wake, decreasing velocity gradients and shear. Eventually, the TKE profiles between the impingement and non-impingement regions approach each other as the porous matrix filters the impact of the macroscale vortex.

Taken together, these results provide strong support for the pore-scale prevalence of turbulence even in the presence of externally forced macroscale vortices. The porous matrix acts as a robust geometric filter that erases the upstream turbulent history and enforces a rapid transition to pore-scale-dominated dynamics. These findings have implications for the design of porous materials used in flow control, filtration, heat transfer, and drag-modification applications, where external turbulence is unavoidable. More broadly, they demonstrate that the interaction between external coherent structures and porous media must be understood through the redistribution of turbulent kinetic energy across scales, rather than through the survival of macroscale vortices themselves.

Future work should aim to confirm these findings in 3D to investigate the impact of inherently 3D mechanisms such as vortex stretching and the energy cascade. It may also explore the effects of varying obstacle shapes and scale separation on vortex behavior, as well as the impact of random solid-obstacle arrangements. Assessing how stochasticity of the porous medium alters penetration depth, interfacial production, turbulence and pressure diffusion, and dissipation across successive pore layers will clarify the role of geometric variability on turbulence transport in porous media. It may also explore the impact of the length of the porous matrix on the TKE.

**Acknowledgements.** The authors acknowledge the computing resources provided by North Carolina State University High Performance Computing Services Core Facility (RRID:SCR_022168). AVK acknowledges the support of the Alexander von Humboldt Foundation through the Humboldt Research Award.

**Funding.** This research was funded by the National Science Foundation under award CBET-2042834

**Declarations of interests.** The authors report no conflict of interest.

**Data availability statement.** The data that support the findings of this study are available from the corresponding author upon reasonable request.

**Author ORCIDs.** V. Srikanth, https://orcid.org/0000-0002-2521-3323;

A. V. Kuznetsov, https://orcid.org/0000-0002-2692-6907

# Scale Collapse of Vortices at Porous/Fluid Interfaces

Justin Courter, Vishal Srikanth, Thibaut Kemayo, and Andrey V. Kuznetsov

Department of Mechanical and Aerospace Engineering, North Carolina State University, Raleigh, NC 27695, USA

## Supplemental Materials

### S.1 Flow-state diagnostics and interpretation of pore-scale unsteadiness

The bluff body cases are run at $Re_D = 10{,}000$, based on the bluff body side length $D$ and the inlet velocity 1 m/s, so that the incoming wake contains large-amplitude macroscale vortical structures before reaching the porous layer. However, Reynolds number alone is not used as evidence that the flow inside the porous matrix is turbulent. This distinction is essential because the no-bluff body case has laminar inflow upstream of the porous layer. Therefore, the flow state inside the porous matrix is diagnosed directly from the resolved velocity field.

It is important to establish that the velocity fluctuations associated with the impinging vortices downstream of the bluff body in the clear-fluid region are not solely the unsteady velocity fluctuations associated with deterministic vortex shedding but also contain the broadband fluctuations consistent with transitional/turbulent shear-layer dynamics. Bai & Alam [1] classified $Re > 1.0 \times 10^3$ as belonging to the shear layer transition II regime where transition occurs upstream of the trailing corner of a square bluff body; the flow conditions at $Re = 10{,}000$ used within this study lie well within this regime. To further assess the spectral character of the flow in the clear-fluid bluff body wake and within the porous matrix, the Fast Fourier Transform (FFT) was computed along the dotted lines in Figure 1 in the main text; from left to right in the figure, these locations are: (i) midway between the bluff body and porous matrix, (ii) at the clear fluid-porous interface, and (iii) within the porous matrix. The FFT is then converted to an energy-density spectrum as a function of the Strouhal number ($E_{uu}(St)$) as illustrated in Figure S1 for the case $\phi = 0.80$.

The resulting spectra were used to distinguish the narrowband vortex-shedding behavior associated with the bluff body length scale from the broadband spectral content characteristic of transitional/turbulent velocity fluctuations. The clear-fluid spectrum contains broadband energy over a range of Strouhal numbers, indicating that the incoming wake is not purely periodic. The porous matrix energy-density spectrum is shifted upwards relative to that in the clear-fluid region, indicating that the porous matrix modifies the distribution and persistence of $E_{uu}$ across different $St$ and exhibits a broadband energy decay over a wide range of frequencies. The flow conditions, geometry configuration, and mesh sizing upstream of the porous matrix are identical for all cases with a large bluff body near the inlet, implying that $E_{uu}(St)$ decays similarly for all porosities in the clear-fluid region; however, the energy-density spectrum within the porous matrix will vary with porosity. Still, all FFTs of the fluctuating components of velocity within the porous matrix contain a wide band of frequencies, regardless of porosity. Therefore, it may be concluded that all externally generated impinging vortices are transitional/turbulent and stochastic velocity fluctuations develop within the porous matrix. A similar analysis yields the same conclusion in the region of the porous matrix not directly downstream of the bluff body in which no vortex impingement occurs.

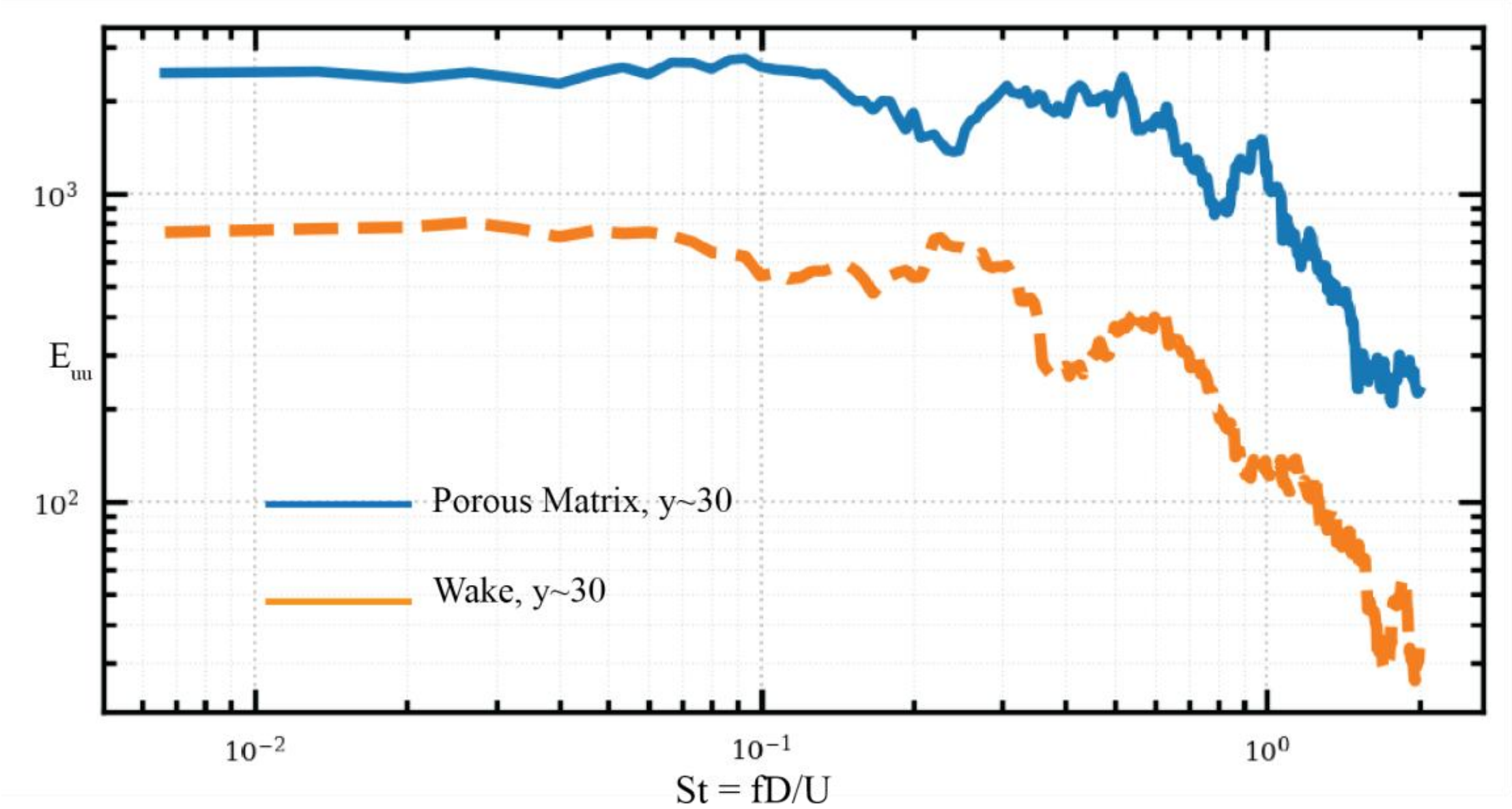

Figure S1: Energy-density spectrum as a function of Strouhal number in the clear-fluid, impingement region upstream of the porous matrix (orange) and within the porous matrix (blue). The clear-fluid wake region energy indicates that the impinging flow includes stochastic velocity fluctuations consistent with transitional/turbulent regimes. Note that $E_{uu}$ at the fluid/porous interface is not included because the spectrum appears very similar to the wake region.

### S.2 Justification for performing 2D rather than 3D DNS

A 3D DNS of the present configuration was determined to be computationally prohibitive. The present 2D simulation for $\phi$=0.80 used $N_{2D} = 1.14 \times 10^6$ cells, $Re = 10{,}000$, $\Delta t = 0.05\ s$, and required 6.46 days on 40 cores to reach a statistically stationary state at $t = 6{,}800\ s$; here, $N_{2D}$ is the total number of cells. A minimally resolved spanwise extension requires resolution down to $\Delta x = \Delta y = \Delta z = 0.0035$ which is the first layer height at the solid obstacles in the porous matrix where $\Delta x$ and $\Delta y$ are the porous-grid resolutions used in the present simulations and $\Delta z$ would be the spanwise, 3D extension, first layer height. Taking the arithmetic mean of the global mesh size and first layer mesh resolution yields a representative cell size in the porous matrix of $\Delta z_{eff} = \frac{0.0035+0.1}{2} = 0.05175$. Using an expansion in the spanwise direction of $L_z = 6D$ gives $N_z = L_z/\Delta z_{eff} \approx 1{,}159$ additional spanwise planes, resulting in a total mesh size of $N_{3D} = N_{2D} \times N_z = 1.32 \times 10^9$ cells. Based on the 2D wall-clock run time of 6.46 days utilizing 40 cores, the run cost is estimated as $6{,}201.6\ core \times hr$ which would correspond to $7.19 \times 10^6\ core \times hr$ in 3D or roughly 19.9 years of wall-clock time on the same 40-core allocation. A more conservative allocation of $L_z = 12D$ would double these estimates; therefore, a fully resolved 3D DNS is beyond the computational scope of the present study.

To resolve the microscale turbulent structures at a feasible computational cost, the geometry is assumed to be 2D, similar to prior studies that employed 2D simulation models to resolve turbulence and develop fundamental scaling behaviors [2–5]. The objective of this study is to develop a theoretical, qualitative understanding of how macroscale vortices interact with the microscale pores of a porous layer. To this end, we employ the 2D approximation as a theoretical limit, isolating the dynamics of coherent structures from stochastic background turbulence. We acknowledge that 2D DNS inherently neglects the vortex stretching mechanism essential for the forward energy cascade in 3D flows. While this omission prevents the prediction of realistic vortex breakdown, it enables a rigorous characterization of the fundamental shedding modes and their interaction with pore boundaries. It also enables simulations performed over a wide range of porosities, at high $Re_D$, and with a large computational domain.

Boffetta & Ecke [6] note that 2D turbulence results in enstrophy conservation in the inviscid limit because of the loss of the vortex stretching mechanism when moving from 3D to 2D. This results in a direct enstrophy cascade to small scales and an inverse kinetic energy cascade to large scales. Consequently, we anticipate a prolonged survival of large-scale eddies in 2D turbulence compared to 3D because of reduced dissipation in the absence of the vortex stretching mechanism. Precise quantitative predictions of drag are outside the scope of this work; instead, turbulence statistics are interpreted via the Kraichnan-Leith-Batchelor theory of 2D turbulence. In this framework, TKE values represent the energy stored in the large-scale coherent vortices. TKE production serves as a theoretical maximum for energy transfer from the mean shear, free from spanwise losses. Finally, the dissipation rate is interpreted as enstrophy dissipation, representing the viscous smoothing of vorticity gradients rather than kinetic energy loss. This framework establishes a limiting scenario for vortex persistence. If macroscale structures are attenuated in this 2D limit, they would very likely be destroyed in a fully 3D environment. Further discussion and benchmarking against previous studies is provided in §S.5.

### S.3 Numerical procedure, sampling, and TKE metrics

The DNS simulations are performed using ANSYS Fluent 2025 R1 with a finite-volume formulation. DNS simulations do not require turbulence modeling of any kind, so they are the only fully resolved numerical models that can answer whether macroscopic turbulence can exist in porous matrices [7], [8], [9]. The unsteady incompressible Navier-Stokes equations are solved directly on the explicitly resolved 2D geometry. Pressure-velocity coupling is treated with the coupled algorithm. Velocity and pressure are stored at staggered locations on the computational cells to minimize interpolation error (PRESTO! Scheme). The convective term of the momentum equation is discretized using the Quadratic Upwind Interpolation for Convective Kinematics (QUICK) scheme, whereas the diffusive term is discretized with the second order central differencing scheme. Time integration is performed with a second-order implicit scheme. The simulation is first advanced until a statistical stationary state is reached, which we monitored based on the drag coefficient and root-mean-square velocity monitors. After statistical stationarity is reached, we collected turbulence statistical data over the simulation domain until the second order statistics were observed to be converged. The turbulent kinetic energy is evaluated as

$$k = \frac{1}{2}(\langle u'^2 \rangle + \langle v'^2 \rangle) \quad \textbf{(1)}$$

where u' and v' are the temporal velocity fluctuations in the streamwise and transverse directions, respectively. This quantity is used to identify where fluctuation energy is introduced, redistributed, or dissipated within the porous matrix.

The integral length scale of the vortical structures is quantified using a two-point velocity correlation function ($R$) along the transverse direction.

$$R(dy, y_0) = \frac{\langle u'(y_0, t) \times u'(y_0 + dy, t) \rangle_t}{\langle u'^2(y_0, t) \rangle_t} \quad \textbf{(2)}$$

where $y_0$ is the reference point, taken at the midpoint of the transverse sampling line, $\Delta y$ is the transverse separation distance, u' is the fluctuating streamwise velocity, and $\langle\cdot\rangle_t$ denotes time averaging. The width of the correlation function is used to estimate the size of the coherent structures in both the bluff body wake and porous layer, which is supplemented by the TKE to determine whether the incoming macroscale vortices persist inside the matrix or are reduced to pore-scale motions. The TKE transport analysis (see §S.6 below for full expression) is used to interpret the mechanism responsible for this scale reduction. The budget separates the local TKE balance into production, pressure diffusion, viscous diffusion, turbulent transport, and viscous dissipation.

### S.4 Grid quality and temporal resolution

In the present study, our goal is to investigate the macroscale turbulent vortex structures as well as the microscale structures that exist in the porous matrix. The flow properties associated with these structures are utilized to generate the governing terms in the TKE transport equations, the TKE, and the velocity correlation functions. In this section, we show that the grid and temporal resolution used is adequate to achieve these objectives considering the bounding porosities of $\phi = 0.80, 0.95$. Since the grid size in all cases is the same, the adequacy of the grid resolution for the bounding porosities is assumed to be applicable for the other porosities as well. The Reynolds number at the large solid obstacle is maintained at 10,000, and the near wall grid resolution, $y^+$, is held constant such that the non-dimensional near wall grid height is less than 1.

$$y^+ = \frac{y \times \rho \times u_\tau}{\mu} \quad \textbf{(3)}$$

where $\rho$ is the fluid density in kg/m$^3$, $u_\tau$ is the friction velocity in $m/s$, $y$ is the wall-normal height of the first cell in $m$, and $\mu$ is the fluid viscosity in $Pa \times s$. The baseline global element size was increased by factors of 1.50 and 2.00. For each mesh, the solution was advanced until the solution reached a statistically stationary state. The mesh and temporal sensitivity were both evaluated using the time-averaged porous drag coefficient $\overline{C_D}$ which was monitored at 0.05 s intervals after statistical stationarity was reached, and the reported values were averaged over the final statistically stationary interval.

The results in Table S1 show a weak sensitivity to mesh resolution. Changing the number of mesh elements from 1,137,361 to 522,178 by doubling the mesh size resulted in only a 0.40% decrease of $\overline{C_D}$ for $\phi = 0.80$; doubling the mesh size and changing from 965,915 to 598,251 elements resulted in a 0.87% decrease for $\phi = 0.95$.

A second study was conducted to investigate the sensitivity of $\overline{C_D}$ to timestep size. The timestep was increased by a factor of 1.50, changing the timestep from 0.05 s to 0.075 s, and a similar process to that described above was performed. Increasing the timestep size resulted in a change in $\overline{C_D} < 2\%$, indicating that the selection of the timestep size is sufficient for this study.

Therefore, the baseline mesh and timestep resolution are sufficiently resolved for the purposes of this study, while maintaining a reasonable computational cost.

Table S1: Mesh and Timestep Sensitivity Study

| **Porosity** | **Resolution Study** | **Mesh Size Multiplier** | **Global Size** | **Nodes** | **Elements** | **Timestep (s)** | $\overline{C_D}$ | **Percent Error (%), Relative to 1.00x** |
|---|---|---|---|---|---|---|---|---|
| $\phi = 0.80$ | Mesh | 1.00 (Nominal) | 0.10 | 1,150,148 | 1,137,361 | 0.05 (Nominal) | 642.934 | N/A |
| | | 1.50 | 0.15 | 693,824 | 681,686 | 0.05 | 641.217 | 0.267 |
| | | 2.00 | 0.20 | 533,931 | 522,178 | 0.05 | 640.349 | 0.402 |
| | Temporal | 1.00 | 0.10 | 1,150,148 | 1,137,361 | 0.075 | 632.063 | 1.691 |
| $\phi = 0.95$ | Mesh | 1.00 (Nominal) | 0.10 | 978,351 | 965,915 | 0.05 (Nominal) | 353.507 | N/A |
| | | 1.50 | 0.15 | 770,642 | 758,412 | 0.05 | 355.417 | 0.540 |
| | | 2.00 | 0.20 | 610,091 | 598,251 | 0.05 | 348.915 | 0.870 |
| | Temporal | 1.00 | 0.10 | 978,351 | 965,915 | 0.075 | 350.432 | 1.299 |

## S.5 Two-dimensional framework and benchmark interpretation

After the mesh and temporal resolution studies, the two-dimensional pore-resolved formulation was assessed against the pore-resolved DNS data of Chu *et al.* [10] for flow through a periodic array of square cylinders. The purpose of this comparison is not to establish quantitative equivalence with three-dimensional turbulence, but to determine whether the present model reproduces the leading pore-scale hydrodynamic features relevant to the present analysis. Accordingly, the benchmark is restricted to the mean streamwise velocity and the streamwise velocity-fluctuation intensity, which are the two quantities most directly connected to vortex redistribution and turbulent kinetic-energy transport in this study.

Figure S2 compares the present profiles with the DNS data of Chu *et al.* [10] at five streamwise locations, $x = 1$, 1.5, 2, 2.5, and 3. The mean streamwise velocity is captured with reasonable qualitative accuracy, including the acceleration through the pore openings and the subsequent redistribution induced by the neighboring solid obstacles. The fluctuation-intensity profiles also reproduce the main spatial organization of the DNS data, particularly the concentration of velocity fluctuations within the inter-obstacle region. Differences in magnitude are expected because the present simulations are two-dimensional and therefore do not include spanwise instabilities, vortex stretching, or the associated three-dimensional energy cascade. The benchmark therefore supports the use of the present formulation as a qualitative pore-resolved surrogate for analyzing wake-induced fluctuation redistribution in the porous matrix, while the reported fluctuation magnitudes should not be interpreted as quantitatively equivalent to fully three-dimensional DNS.

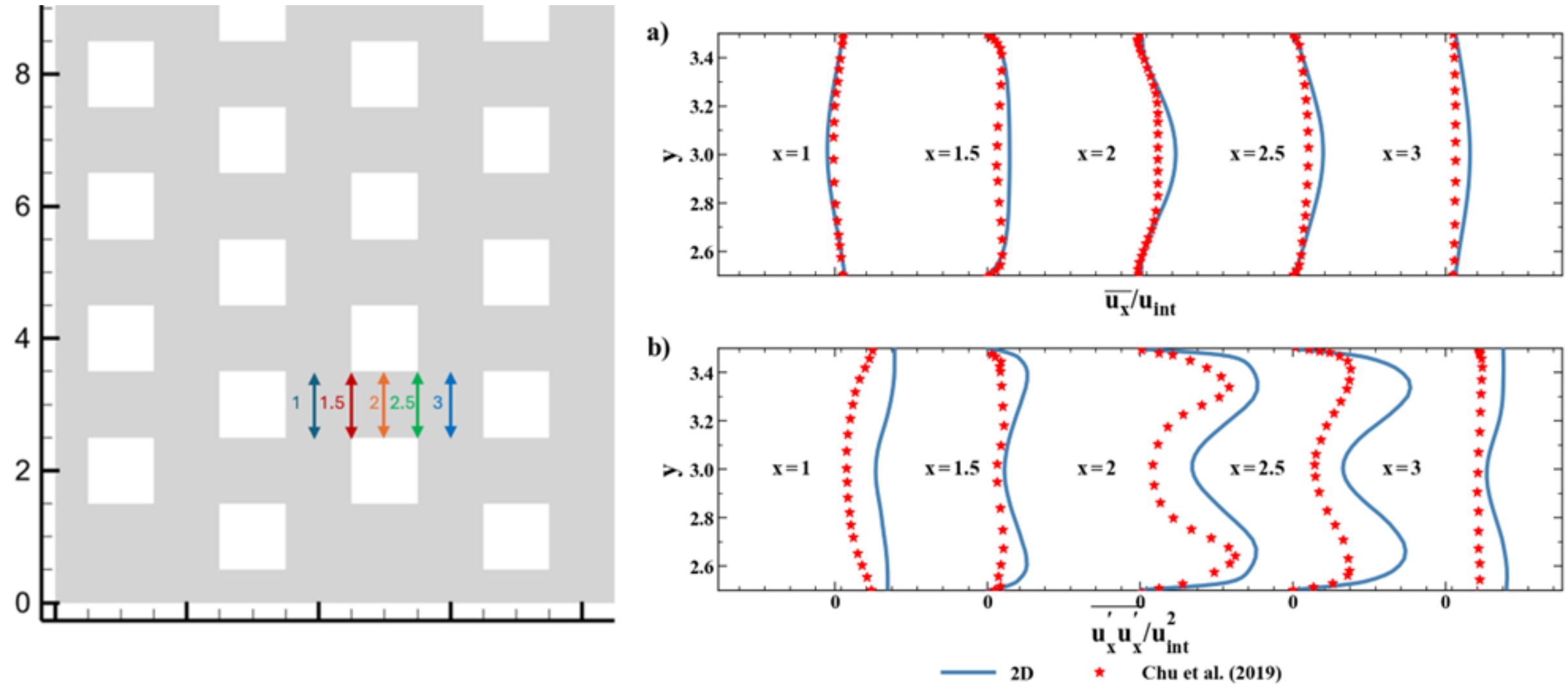

Figure S2: Benchmark comparison between the present two-dimensional pore-resolved simulations and the three-dimensional DNS data of Chu *et al.* [10] for flow through a periodic square-cylinder array. The left panel shows the pore geometry and the profile-sampling locations. The right panels compare (a) the mean streamwise velocity, $\overline{u_x}/u_{int}$, and (b) the streamwise velocity-fluctuation intensity, $\overline{u'_x u'_x}/u^2_{int}$, at $x = 1$, 1.5, 2, 2.5, and 3. The comparison shows that the present model captures the main pore-scale hydrodynamic structure, while deviations in fluctuation intensity reflect the absence of three-dimensional turbulence mechanisms in the two-dimensional formulation.

## S.6 Turbulent kinetic energy transport equation

The volume-averaged TKE transport equation is generated by simplifying the 3D TKE transport equation to 2D based on Section 4.2 in [11]. It is rewritten as to allow calculation of the TKE budget terms using mean rather than fluctuating quantities so that each term can be calculated from sampled data in ANSYS.

Production Term

$$P = -\left[\left(\overline{u^2} - \bar{u}^2\right)\frac{\partial \bar{u}}{\partial x} + \left(\overline{uv} - \bar{u}\bar{v}\right)\frac{\partial \bar{u}}{\partial y} + \left(\overline{vu} - \bar{v}\bar{u}\right)\frac{\partial \bar{v}}{\partial x} + \left(\overline{v^2} - \bar{v}^2\right)\frac{\partial \bar{v}}{\partial y}\right] \quad \textbf{(4)}$$

Pressure Diffusion Term

$$D_p = -\frac{1}{\rho}\left[\frac{\partial}{\partial x}\left(\overline{up} - \bar{u}\bar{p}\right) + \frac{\partial}{\partial y}\left(\overline{vp} - \bar{v}\bar{p}\right)\right] \quad \textbf{(5)}$$

Turbulent Transport Term

$$\begin{aligned} T = -\frac{1}{2}\Bigg[\frac{\partial}{\partial x}\Big(\overline{u^3} + \overline{v^2u} - \big(\overline{u^2}\bar{v} + \overline{v^2}\bar{u}\big) - 2\big(\bar{u}\overline{u^2} + \bar{v}\overline{vu}\big) + 2\big(\bar{u}^3 + \bar{v}^2\bar{u}\big)\Big) \\ + \frac{\partial}{\partial y}\Big(\overline{v^3} + \overline{u^2v} - \big(\overline{u^2}\bar{u} + \overline{v^2}\bar{v}\big) - 2\big(\bar{v}\overline{v^2} + \bar{u}\overline{uv}\big) \\ + 2\big(\bar{v}^3 + \bar{u}^2\bar{v}\big)\Big)\Bigg] \end{aligned} \quad \textbf{(6)}$$

Viscous Diffusion Term

$$D_\nu = \frac{\nu}{2}\left[\frac{\partial^2}{\partial x^2}\left(\left(\overline{u^2} - \bar{u}^2\right) + \left(\overline{v^2} - \bar{v}^2\right)\right) + \frac{\partial^2}{\partial y^2}\left(\left(\overline{u^2} - \bar{u}^2\right) + \left(\overline{v^2} - \bar{v}^2\right)\right)\right] \quad \textbf{(7)}$$

Dissipation Term

$$\varepsilon = \nu\left[\overline{\left(\frac{\partial u}{\partial x}\right)^2} + \overline{\left(\frac{\partial u}{\partial y}\right)^2} + \overline{\left(\frac{\partial v}{\partial x}\right)^2} + \overline{\left(\frac{\partial v}{\partial y}\right)^2} - \left[\left(\frac{\partial \bar{u}}{\partial x}\right)^2 + \left(\frac{\partial \bar{u}}{\partial y}\right)^2 + \left(\frac{\partial \bar{v}}{\partial x}\right)^2 + \left(\frac{\partial \bar{v}}{\partial y}\right)^2\right]\right] \quad \textbf{(8)}$$

## S.7 Integral length scales

Table S2 provides the integral length scales associated with the correlation functions in Figure 4 in the main text. The integral length scale is the area underneath the correlation function prior to its first zero crossing.

Table S2: Integral Length Scales (ILS)

| **Porosity ($\phi$)** | **0.80** | **0.85** | **0.90** | **0.95** |
|---|---|---|---|---|
| Porous Interface ILS (m) | 1.54 | 1.68 | 1.66 | 1.66 |
| Middle of the Porous Matrix ILS (m) | 0.10 | 0.17 | 0.21 | 0.24 |
| Middle of the Porous Matrix in the Non-Impingement Region ILS (m) | 0.10 | 0.16 | 0.22 | 0.24 |

## S.8 Supplementary figures

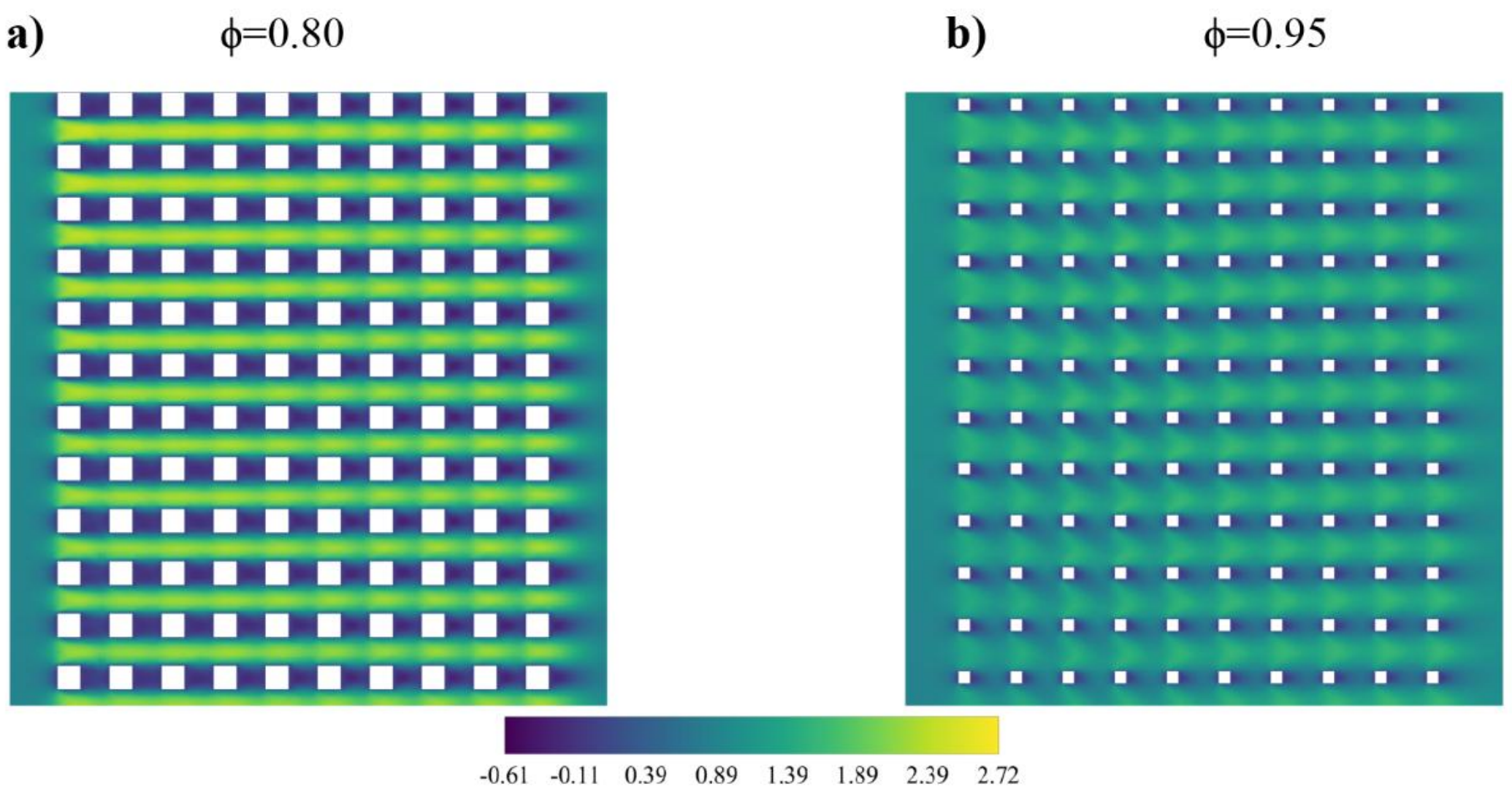

Figure S3: Mean x-velocity contours for $\phi = 0.80$ and $\phi = 0.95$ within the porous matrix.

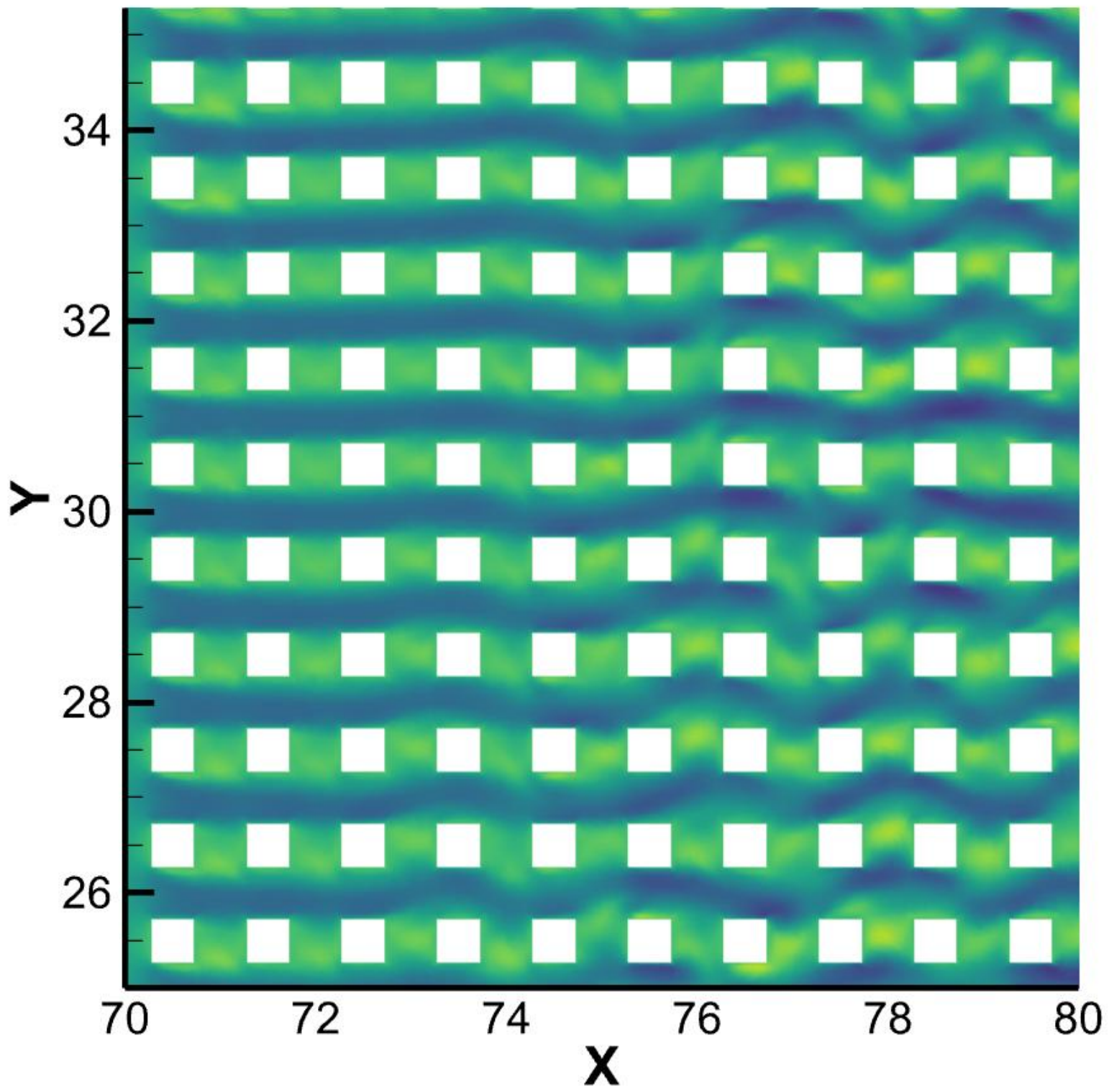

Figure S4: Instantaneous x-velocity contours for $\phi = 0.80$. The downstream growth of TKE is eventually arrested by a mechanism of wake-channel blending wherein microscale eddies formed in the obstacle wakes increasingly oscillate into the adjacent high-speed interstitial channels due to secondary flow instabilities. This can be seen in the decrease and eventual levelling out of the mean x-velocity beginning around 5 obstacles into the porous matrix. This impact can be seen in both the vortex impinged and impingement-free regions of the porous matrix.

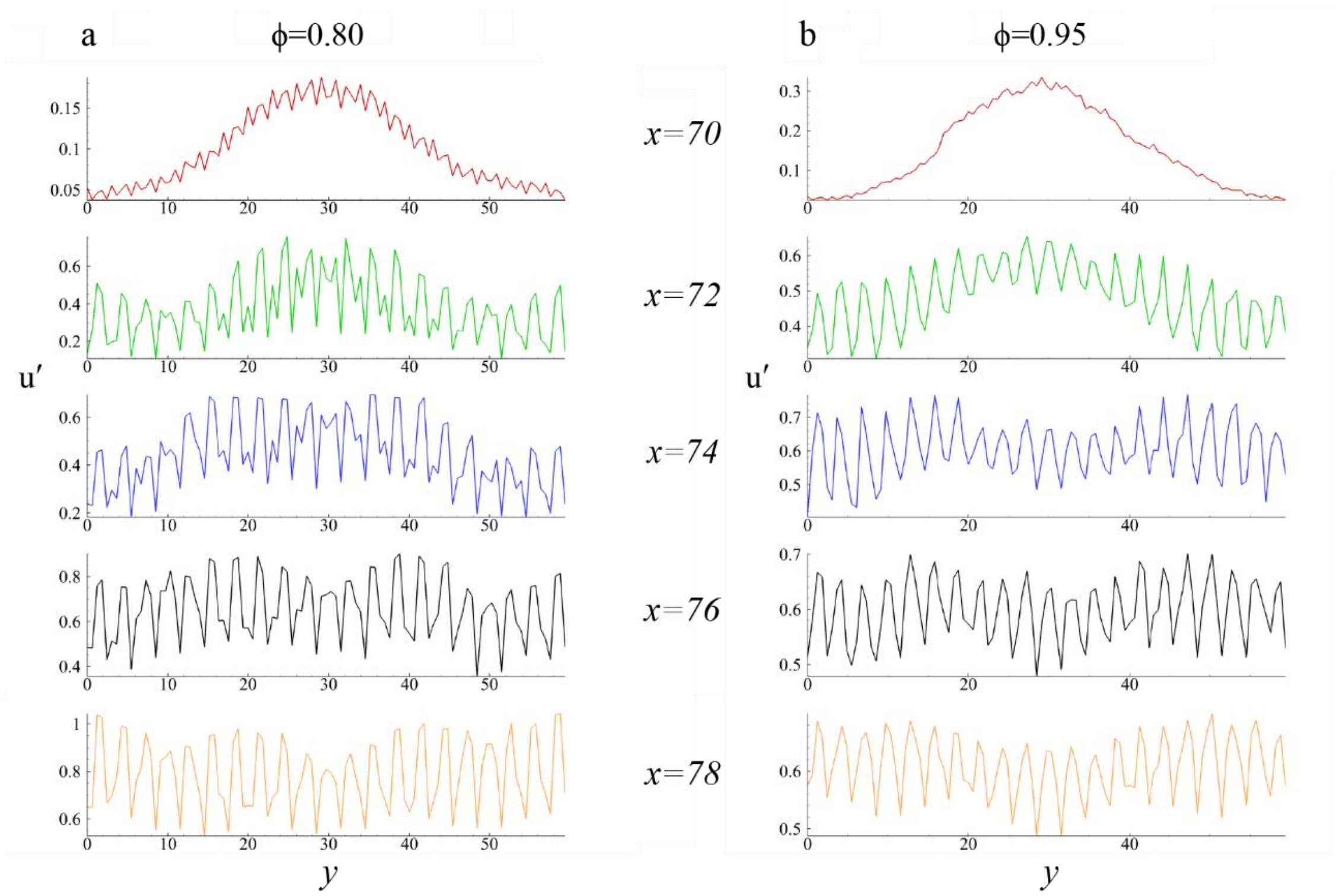

Figure S5: Evolution of the streamwise velocity fluctuations, $u'$, through the porous matrix for (a) $\phi = 0.80$ and (b) $\phi = 0.95$. In both cases, the macroscale vortical structure diminishes within approximately 2–5 obstacle rows. A persistent momentum deficit is evident near $y \approx 30$ (beginning at $x \approx 76$ for $\phi = 0.80$ and $x \approx 74$ for $\phi = 0.95$), directly downstream of the bluff body, where the streamwise velocity fluctuations in the vortex-impinged region remains lower than in the neighboring impingement-free regions.

## S.9 References for the Supplemental Material